\documentclass[a4paper]{jpconf}
\usepackage{graphicx}
\usepackage{amsmath}
\newcommand{\lyxmathsym}[1]{\ifmmode\begingroup\def\b@ld{bold}
  \text{\ifx\math@version\b@ld\bfseries\fi#1}\endgroup\else#1\fi}

\begin{document}
\title{Simplified bond-hyperpolarizability model of second-harmonic-generation in Si(111): theory and experiment}

\author{Hendradi Hardhienata\footnote{on leave from Theoretical Physics Division, Bogor Agricultural University,  Jl. Meranti S, Darmaga, Indonesia}}
\address{Center for surface and nanoanalytics (ZONA), Johannes Kepler University, Altenbergerstr. 69, 4040 Linz, Austria}

\author{Andrii Prylepa}
\address{Christian Doppler laboratory for microscopic and spectroscopic material characterization, Johannes Kepler University, Altenbergerstr. 69, 4040 Linz, Austria}
\address{Center for surface and nanoanalytics (ZONA), Johannes Kepler University, Altenbergerstr. 69, 4040 Linz, Austria}

\author{David Stifter}
\address{Christian Doppler laboratory for microscopic and spectroscopic material characterization, Johannes Kepler University, Altenbergerstr. 69, 4040 Linz, Austria}
\address{Center for surface and nanoanalytics (ZONA), Johannes Kepler University, Altenbergerstr. 69, 4040 Linz, Austria}

\author{Kurt Hingerl}
\address{Center for surface and nanoanalytics (ZONA), Johannes Kepler University, Altenbergerstr. 69, 4040 Linz, Austria}

\ead{kurt.hingerl@jku.at}

\begin{abstract}
Second-harmonic-generation (SHG) in centrosymmetric material such as Si(111) is usually understood either from the phenomenology theory or more recently using the simplified bond-hyperpolarizability model (SBHM)  [G.  D.  Powell,  J.  F.  Wang,  and  D.  E.  Aspnes,  Phys.  Rev.  B  65, 205320/1 (2002)].  Although SBHM is derived from a classical point of view, it has the advantage over the former that  it gives especially for lower symmetry systems a clear physical picture and a more efficient explanation of how nonlinearity is generated.  In this paper we provide a step-by-step description of the SBHM in Si(111) for the linear and second harmonic case. We present a  SHG experiment of Si(111) and show how it can be modelled by summing up the contribution of the fields produced by the anharmonic motion of the charges along the bonds.
\end{abstract}

\section{Introduction}
Nonlinear optics is one of the most versatile tools to investigate surface properties, especially for a material class which posseses inversion symmetry. Serious theoretical efforts to explain nonlinearity in such systems has significantly progressed especially since the pioneering work of Bloembergen $et. al.$ [1]. A nonlinear optical probe has several advantages over other surface probes because the material damage and contamination associated with charged particle beams is eliminated, all pressure ranges are accessible,  insulators can be studied without the problem of charging effects,  and buried interfaces are accessible owing to the large penetration depth of the optical radiation [2]. In centrosymmetric material such as silicon (Si) the bulk and surface nonlinear contributions are comparable [3] and changes in the surface properties such as surface deposition will significantly alter the nonlinear intensity profile [4].
 
Because centrosymmetric materials is important to present day technology (e.g. silicon semiconductor, silicon thin film sensors) understanding the surface property and the physical mechanism behind is critical, with much effort both experimental and theoretical having been invented. Perhaps, the most notable development  in understanding the creation of an outgoing $2\omega$ wave or second-harmonic-generation (SHG) from centrosymmetric material surfaces and interfaces in the viewpoint of the phenomenological theory was performed by the group of Sipe $et. al.$ [5] and further by Luepke $et. al.$  [6]. The later group, calculated the nonlinear susceptibility for vicinal Si(111) by fitting a fourier series to reproduce the four polarization SHG intensity ($p$-incoming $p$-outgoing, $ps$, $sp$, and $ss$) obtained from their experiments. Unfortunately, already for a single atom with a relatively low symmetry  such as silicon, the phenomenological theory requires many -seemingly non independent- variables to be fitted thus blurring the physical insights and mechanism of SHG inside the material.

As a response to this inefficiency,  Aspnes $et. al.$ [7] developed an ingenious method termed the simplified bond-hyperpolarizability model (SBHM) based on the somehow forgotten classical Ewald-Oseen extinction theory [8,9]. The theorem states  that the electric field transmission and reflection can be obtained by direct superposition of dipoles inside the matherial rather than from macroscopic boundary calculation [10]. SBHM extends this classical view to cope with nonlinear optics by assuming that the SHG signal originates from the anharmonic motions of the charges along the bonds. Suprisingly, if the nonlinear field is far smaller than the linear field  the calculation turns out to be much simpler  because it does not lead to into an integral equation hence requiring self consistency check [11]. In this case it is possible to differentiate between the incoming driving fundamental field ($\omega$) and the outgoing SHG field ($2\omega$). As a result, the steps to obtain the far field contribution from the anharmonic bond radiation can be performed independently. Moreover, experimental data, e.g.Luepke's Si(111) experiment, can be very well fitted using SBHM by only a handful parameters rather than the cumbersome fitting of various fresnel parameters using phenomenology theory.

\section{Basic Theory}
An excellent discussion of how to calculate the far field intensity for centrosymmtric material such as Si(111) and Si(001) by assuming charge motion along their bonds already exists for vicinal Si sytems [6,12] and the interested reader is therefore suggested to refer to these papers for further information. However, for the sake of clarity, we will -although only briefly- show how SHG can be understood in the framework of the SBHM.

In classical electrodynamics one can obtain the  far field expression as a result of dipole (or oscillating charge) radiation. The vector potential  $\vec{A}(\vec{r},t)$, which satisfy the Lorenz gauge [13]:
\begin{equation}
\nabla\cdot\vec{A}=-\frac{1}{c}\frac{\partial\phi}{\partial t}\end{equation}
can be expressed in the form:
\begin{equation}
\vec{A}(\vec{r},t)=\frac{\mu_{0}}{4\pi}\int dt'd^{3}dt'\vec{r'}\frac{\vec{j}(\vec{r'},t'-\frac{\left|\vec{r}-\vec{r'}\right|}{c}-t)}{\left|\vec{r}-\vec{r'}\right|}\end{equation}
Assuming that the localized current  $ j(r)$ and charge density $\rho (r)$ as well as the vector potential $\vec{A}(\vec{r},t)$ can be written in harmonic form we can write eq.(2)  in the far field approximation as:
\begin{equation}
\vec{A}(\vec{r'},t)=\frac{\mu_{0}}{4\pi}\frac{e^{ikr}}{r}\int d^{3}\vec{r'}\vec{j}(\vec{r'})e^{-i\omega t}=-\frac{\mu_{0}}{4\pi}\frac{e^{ikr}}{r}\int d^{3}\vec{r'}\vec{r'}\left|\nabla\cdot\vec{j}(\vec{r'})\right|\end{equation}
Using the continuity equation:
\begin{equation}
\nabla\cdot\vec{j}(\vec{r'})=i\omega\rho(\vec{r})\end{equation}
we have for the spatial dependence vector potential the expression:
\begin{equation}
\vec{A}(\vec{r})=-\frac{i\omega\mu_{0}}{4\pi}\frac{e^{ikr}}{r}\vec{p}\end{equation} 
where $\vec{p}$ is the dipole moment:
\begin{equation}
\vec{p}=\int d^{3}\vec{r'}\vec{r'}\rho(\vec{r})\end{equation}
The expressions for the fields according to the defined vector potential are:
\begin{equation}
\vec{E}(\vec{r})=\sqrt{\frac{\mu_{0}}{\epsilon_{0}}}\frac{i}{k}\nabla\times\vec{H}(\vec{r})\end{equation}
\begin{equation}
\vec{H}(\vec{r})=\frac{1}{\mu_{0}}\nabla\times\vec{A(}\vec{r})\end{equation}
The far field can thus be calculated by inserting eq.(5) in eq.(8) and then inserting eq.(8) in eq.(7). With the help of the identity $(\hat{n}\times \vec{p})\times\hat{n}=\vec{p}-(\hat{n}\hat{n})\cdot\vec{p}$  we have:
\begin{equation}
\vec{E}(\vec{r})=\frac{ke^{ikr}}{4\pi\epsilon_{0}r}\left[(\overline{I}-\hat{n}\hat{n})\cdot\vec{p}\right]\end{equation}
here $\overline{I}$ is the identity tensor. The next step will be finding an expression for the classical microscopic dipole per unit volume (polarization) $\vec{p}$.   In this work we will use eq. (9) to calculate the reflected intensity of the linear and second-harmonic-generation.  

\section{ SBHM  for Si(111)}
In the Maxwell classical approach all the microscopic details in the inhomogenous nonlinear wave equation are included and hidden in the electric susceptibility. In the classical optic approach this susceptibilty can be estimated by constructing a Lorentz oscillator model with damping to obtain the nonlinear susceptibility terms [14]. It is therefore not suprising -but ingenious- to use a similar approach microscopically, by starting from a force equation to find expressions for the polarization. Furthermore it has been shown that even with the simplification that the nonlinear polarization source only occur along the bond direction, the second-harmonic far field from the anharmonic charge radiation is in very good agreement with experimental result for vicinal Si(111). Here, we briefly follow the approach given by Ref. [7].

\begin{figure}[hpbd]
\begin{center}
\includegraphics[width=17cm]{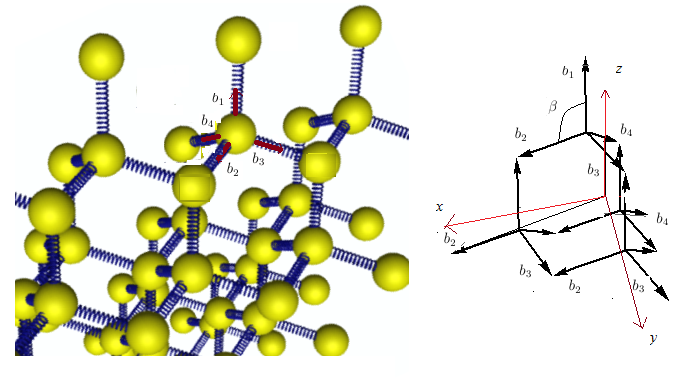}
\end{center}
  \caption{Sketch of Si lattice ((111) surface) and definition of the bond direction in cartesian coordinates.}
  \label{fig:potential}
\end{figure}

For a single Si(111) atom there are four bonds which we denote by $\vec{b}_{1}, \vec{b}_{2}, \vec{b}_{3}$ and $\vec{b}_{4}$. In the bulk, the orientation of the bonds can be expressed in cartesian coordinates as:
\[
b_{1}=\left(\begin{array}{c}
0\\
0\\
1\end{array}\right)\;\quad b_{2}=\left(\begin{array}{c}
\cos\left(\beta-\frac{\pi}{2}\right)\\
0\\
-\sin(\beta-\frac{\pi}{2})\end{array}\right)\]
\begin{equation}
b_{3}=\left(\begin{array}{c}
-\cos\left(\beta-\frac{\pi}{2}\right)\sin(\frac{\pi}{6})\\
\cos\left(\beta-\frac{\pi}{2}\right)\cos(\frac{\pi}{6})\\
-\sin(\beta-\frac{\pi}{2})\end{array}\right)\; b_{4}=\left(\begin{array}{c}
-\cos\left(\beta-\frac{\pi}{2}\right)\sin(\frac{\pi}{6})\\
-\cos\left(\beta-\frac{\pi}{2}\right)\cos(\frac{\pi}{6})\\
-\sin(\beta-\frac{\pi}{2})\end{array}\right)\end{equation}
with $\beta=109.47^\circ$ is the angle between two bonds. Fig.1 depicts a graphical image of the bond direction and the definition of the coordinate.  Note that the four Si(111) bonds are drawn as the red lines. 

Although the orientation is valid for the bulk, the even nonlinear harmonic dipole contribution is generally considered small because it is forbidden by inversion symmetry or parity argument. However we believe that the incoming field decay effects due to absorption may break the symmetry and produce dipolar effects even inside the bulk (spatial dispersion). At the interface, a dipole contribution is allowed because of symmetry breaking, but the orientation of the bonds may not be arranged as homogenously as in the bulk. Nonetheless, we follow the assumption in  Ref. [7] that,  based on statistical averaging the majority of the  bond vector at the interface can be treated similar as within the bulk, thus the bond model of the bulk also holds for the interface.

To model the reflected intensity vs rotation along the azimuthal direction (rotation in the $x$-$y$ plane) we allow the bonds as a function of rotation  via:
\begin{equation}
b_{j}(\phi)=b_{j}\cdot R_{\phi}=b_{j}\cdot\left(\begin{array}{ccc}
\cos\phi & -\sin\phi & 0\\
\sin\phi & \cos\phi & 0\\
0 & 0 & 1\end{array}\right)\end{equation}
where the subscript $j$ runs from 1 to 4.

Having defined the bond vectors, we now analyze the motion of the charges. The force equation which describes this anharmonic motion along the bonds can be written in the form [7] taking the equilibrium position as zero:
\begin{equation}
F=q_{j}\vec{E}\hat{b_{j}}e^{-i\omega t}-\kappa_{1}x-\kappa_{2}x^{2}-b\dot{x}=m\ddot{x}\end{equation}
where  $q, m, x$  are the electron charge, mass, and its displacement from equilibrium,  respectively and $\kappa_{1}$ and $\kappa_{2}$ are the harmonic and anharmonic spring constants, and the term $-b\dot{x}$  is the common frictional loss in oscillation. Solving for $\triangle x_{1}$ and $\triangle x_{2}$ by using the assumption that $x$ can be written as $x=x_{0}+\triangle x_{1}e^{-i\omega t}\triangle x_{2}e^{-i2\omega t}$ gives for the lowest order of approximation:
\[
\triangle x_{1}=\frac{\vec{E}\cdot\hat{b_{j}}}{\kappa_{1}-m\omega^{2}-ib\omega}\]
\begin{equation}
\triangle x_{2}=\frac{\kappa_{2}}{\kappa_{1}-4m\omega^{2}-ib2\omega}\left(\frac{\vec{E}\cdot\hat{b_{j}}}{\kappa_{1}-m\omega^{2}-ib\omega}\right)^{2}\end{equation}
therefore the linear polarization produced by each bond $b_{j}$ is
\begin{equation}
p_{1j}=q_{j}\triangle x_{1}=\alpha_{1j}\left(\hat{b}_{j}\cdot\vec{E}\right)\end{equation}
whereas we have for the SHG the nonlinear polarization:
\begin{equation}
p_{2j}=q_{j}\triangle x_{2}=\alpha_{2j}\left(\hat{b}_{j}\cdot E\right)^{2}\end{equation}
where $\alpha_{1}$ and $\alpha_{2}$ are the microscopic polarizability and second order hyperpolarizability given by: 
\[
\alpha_{1j}=\frac{1}{\kappa_{1}-m\omega^{2}-ib\omega}\]
\begin{equation}
\alpha_{2j}=\frac{\kappa_{2}}{\kappa_{1}-4m\omega^{2}-ib2\omega}\left(\frac{1}{\kappa_{1}-m\omega^{2}-ib\omega}\right)^{2}\end{equation}

In Si(111) surface,  $\alpha_{j=1}$ is denoted as the "up" polarizability/hyperpolarizability $\alpha_{u}$ whereas the three remaining have the same value due to symmetry and are called the "down" polarizability/hyperpolarizability $\alpha_{d}$. The total linear and second harmonic polarization produced by all the four bonds in Si(111) considering azimuthal rotation are thus :
\begin{equation}
\vec{P_{1}}=\frac{1}{V}\left(p_{11}+p_{12}+p_{13}+p_{14}\right)=\frac{1}{V}{\sum\limits_{j=1}^{4}}\alpha_{1j}\hat{b}_{j}\left(\hat{b}_{j}(\phi)\cdot\vec{E}\right)\end{equation}
and
\begin{equation}
\vec{P_{2}}=\frac{1}{V}\left(p_{21}+p_{22}+p_{23}+p_{24}\right)=\frac{1}{V}{\sum\limits_{j=1}^{4}}\alpha_{2j}\hat{b}_{j}\left(\hat{b}_{j}(\phi)\cdot\vec{E}\right)^{2}\end{equation}

In the following chapters, we will show how the SBHM can reproduce experimental results of second-harmonic-generation in Si(111).

\section{ SHG Experiment of Si(111)}

The setup for the azimuthal SHG measurements is shown in Fig. 2. As a source of a radiation a compact femtosecond fiber laser system, working at a wavelength of 1560 nm and producing pulses with duration of 86 fs and with a repetition rate of 80 MHz, was used. After a variable attenuator for adjusting the power, a half-wave plate (HWP) determines the state of polarisation for the incident beam, which is then filtered by a band-pass filter (F), transmitting only the fundamental wavelength. The incidence radiation is focused onto the sample by the objective lens L1 with a minimal focal waist down to 10 $\mu$m.

\begin{figure}[htbp]
\begin{center}
\includegraphics[width=15cm]{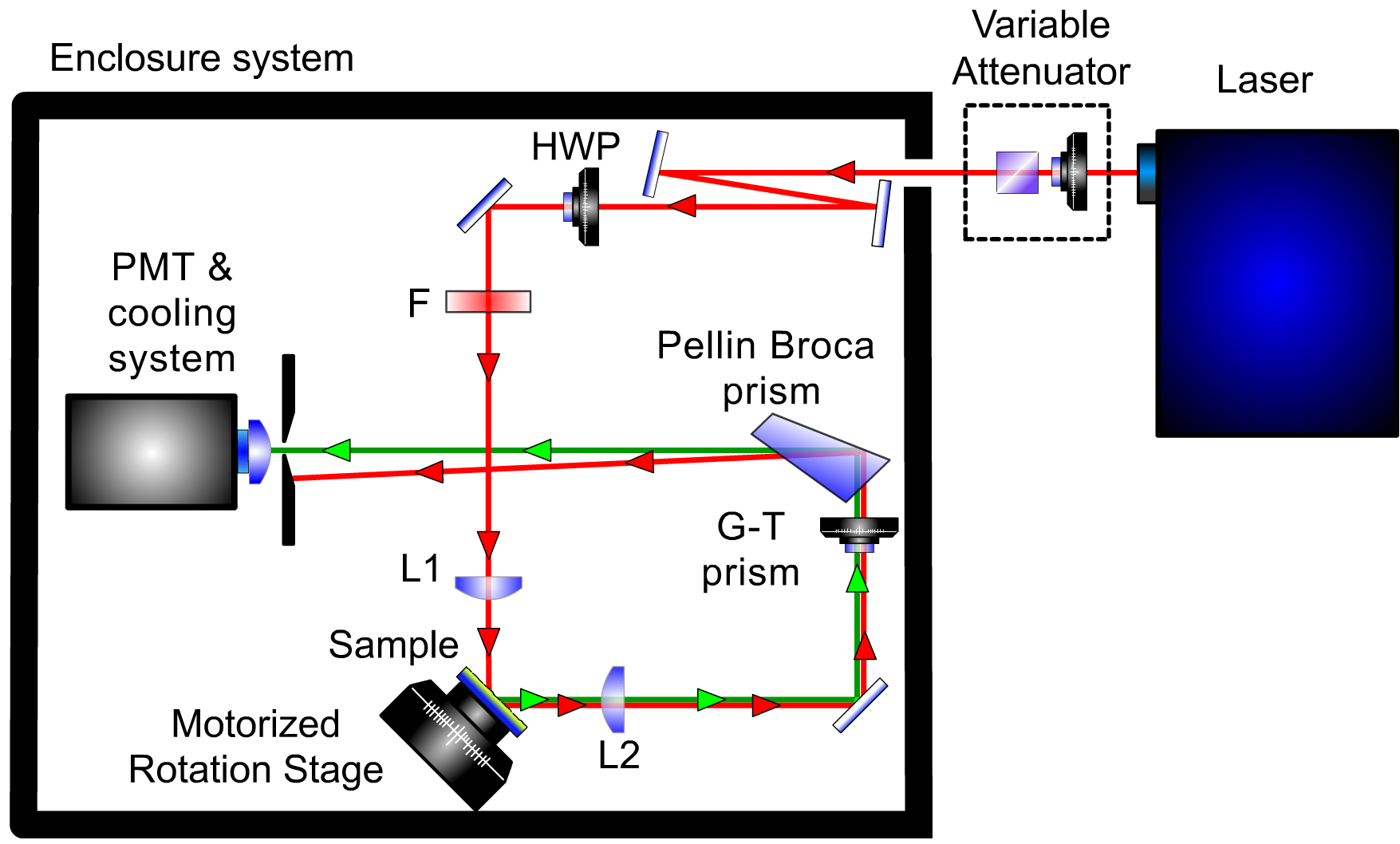}
\end{center}
  \caption{Schematic sketch of the experimental setup for SHG measurements (Abbreviations: HWP: half-wave plate, F:filter, L:lens, G-T:Glan-Taylor polarizing prism, PMT:photomultiplier).}
  \label{fig:potential}
\end{figure}

   The incidence angle for the radiation was set to $45^\circ$. The sample itself is mounted on a motorized rotation stage. SH and fundamental radiation from the sample are collected by the lens L2 and are directed to a rotatable Glan-Taylor calcite polarizer that selects the desired polarisation direction. A Pellin Broca prism and a slit are used for spatially separating the SH radiation, which is focused onto a cooled photomultiplier tube (Hamamatsu R911P). For the acquisition of the SHG signal photon counting was used, with a dark count rate of only 30 photons per minute.

As a sample, a Si(111) wafer with native oxide was used. In order to obtain a detectable SHG signal and not to damage the surface of the sample, the beam was slightly defocused, with the fiber laser operating at full power (350 mW).

 \section{Si(111) SHG experimental results and SBHM simulations }\vspace{2mm}
Second harmonic generation from a Si(111) face was obtained experimentally by plotting the intensity for various azimuthal angles (Si(111) rotated by $360^\circ$ in the $x$-$y$ plane). All four intensity profiles ($pp, ps, sp, ss$) are aquired with the results depicted in Fig. 3(b). To obtain the total field using SBHM we insert expressions for the total polarization in eq. (17) and eq. (18) into the far field formula eq. (9), with $\hat{n}=\widehat{k}$, where $\widehat{k}$ is the outgoing wave vector (the unit vector in the direction of the observer) and is given by $\hat{k}=-\cos \theta_{0} \hat{x}+\sin \theta_{0} \hat{z}$. The linear and second harmonic far field can now be written as:
\begin{equation}
\vec{E}_{\omega g}(r)=\frac{ke^{ikr}}{4\pi\epsilon_{0}rV}\left[(\overline{I}-\hat{k}\hat{k})\cdot{\sum\limits_{j=1}^{4}}\alpha_{1j}\hat{b}_{j}\left(\hat{b}_{j}(\phi)\cdot\vec{E_{g}}\right)\right]\end{equation}
\begin{equation}
\vec{E}_{2\omega g}(r)=\frac{ke^{ikr}}{4\pi\epsilon_{0}rV}\left[(\overline{I}-\hat{k}\hat{k})\cdot{\sum\limits_{j=1}^{4}}\alpha_{2j}\hat{b}_{j}\left(\hat{b}_{j}(\phi)\cdot\vec{E_{g}}\right)^{2}\right]\end{equation}
The term outside the paranthesis includes only constants and because SHG experiments are usually performed in arbitrary units the important terms are those given inside the paranthesis.  Here $g$ is the symbol for the polarization of the incoming wave and can be either $p$ or $s$. 

\begin{figure}[htbp]
\begin{center}
\includegraphics[width=16cm]{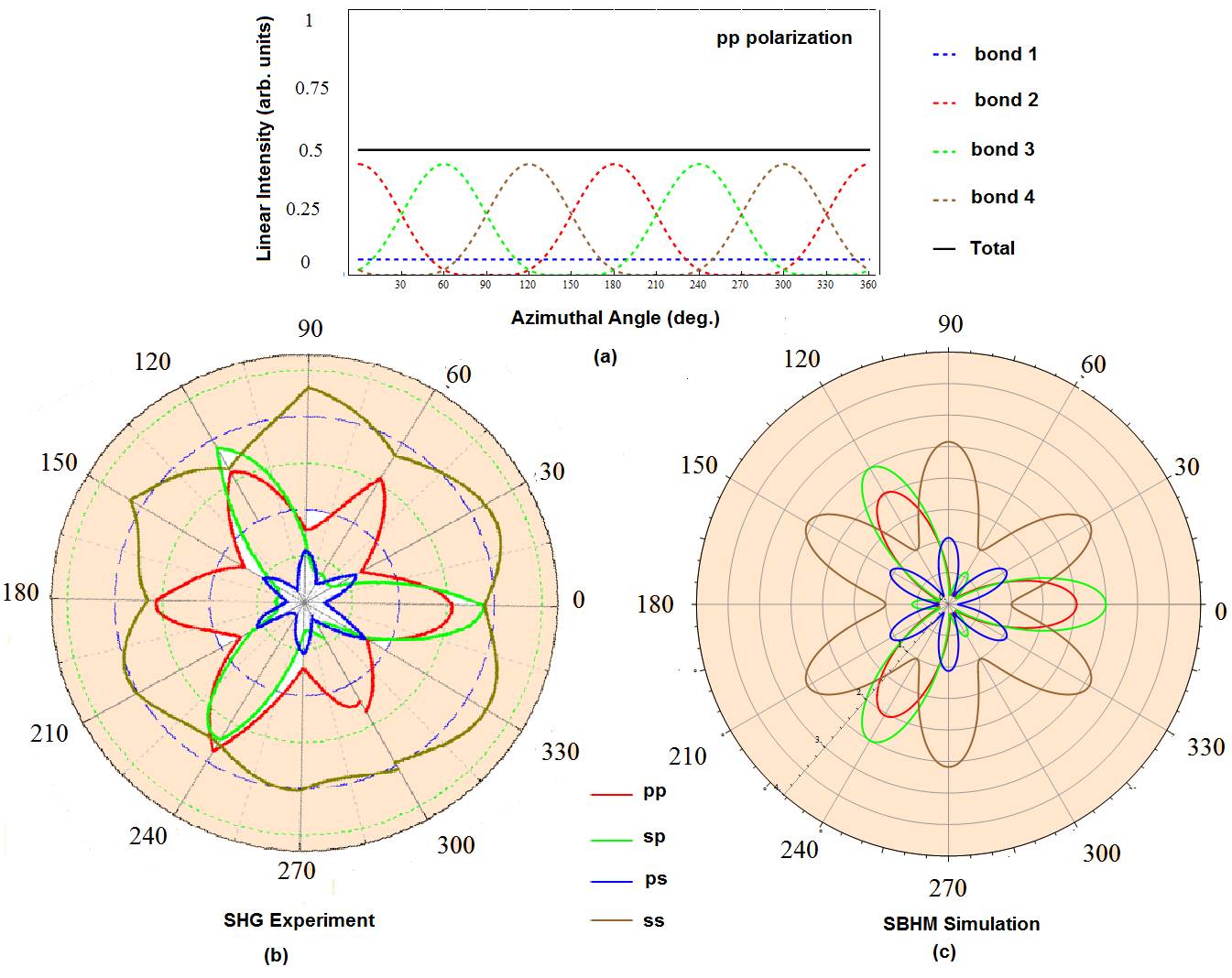}
\end{center}
  \caption{Si(111) linear and SHG  intensity profiles vs azimuthal rotation in the $x$-$y$ plane for various polarizations in arbitrary units of intensity. (a) The total linear pp-intensity yields a constant value independent of the azimutal rotation. (b) SHG intensity obtained from experiment and (c) SHG intensity simulation results using SBHM with $\alpha_{u}$=$\alpha_{d}$=1.}
  \label{fig:potential}
\end{figure}

We next consider first eq. (17) and evaluate the intensity for four polarization direction: $pp, ps, sp$, and $ss$. The first letter in $pp$ denotes the incoming and the second the outgoing wave. The far field from all four bonds in Si(111) for each polarization can be obtained using eq.(10). For the linear case we have:
\begin{equation}
\begin{aligned}
  &\vec{E}_{\omega, pp}=\frac{1}{2}\left(\cos\left(\theta_{\ddot{o}}\right)\hat{x}+\sin\left(\theta_{\ddot{o}}\right)\hat{z}\right)[(-3\text{\ensuremath{\alpha_{d}}}\cos\left(\ensuremath{\theta_{i}}\right)\cos\left(\ensuremath{\theta_{o}}\right)\sin^{2}\left(\beta\right)+3\text{\ensuremath{\alpha_{d}\text{\ensuremath{\cos\left(2\beta\right)}}\sin\left(\ensuremath{\theta_{i}}\right)\sin\left(\ensuremath{\theta_{o}}\right))]}} \\
       &\vec{E}_{\omega, sp}=0  \\
       &\vec{E}_{\omega, sp}=0  \\\label{eq:22}
       & \vec{E}_{\omega, ss}=\frac{3}{2}\alpha_{d}\sin^{2}\left(\beta\right)\hat{y}\\
\end{aligned}
\end{equation}

An interesting result from this calculation is that the total linear fields for all polarization is independent of the azimuthal rotation $\phi$, eventhough the field produced by each of the bond may depend on $\phi$. This can be seen in Fig. 3(a) where we take the $pp$-case as an example. Similar results are obtained for the $ss$-polarization. Therefore, orientation of the bonds due to rotation in the $x$-$y$ plane does not effect the reflected linear intensity.

More interesting results are obtained if we consider the second harmonic case. The total nonlinear polarization for all polarization modes are -in contrary to the linear case- a function of the azimuthal rotation of the crystal $\phi$. Fig. 3(b) shows our experimental result. The $pp, ps$, and $ss$-polarization (red line) showsa 6-fold dependence, with peaks at every $60^\circ$ interval. The $ps$ and $ss$-polarization have the same azimuthal behaviour with  a peak shift of $30^\circ$ for the $pp$-polarization. The $ss$-polarization, notably has a high (DC) offset value and broad peak full-wave-half-maximum (FWHM), but peaks still being detected. Also the height of the $pp$ and $sp$-polarization peaks are quite similar with the sp polarization having a 3-fold symmetry. 

Remarkably, this seemingly complicated experimental result can be well explained using the SBHM, more explicitly by applying  eq. (19) with the result depicted in Fig. 3(c). The formulas for the 4-polarization SHG far field from a Si(111) structure can be seen in Ref. [7]: 
\begin{equation}
\begin{aligned}
&\vec{E}_{2\omega,sp}=\frac{3}{4}\alpha_d\sin^2\beta\left[\cos\theta_o\cos3\phi\sin\beta-2\cos\beta\sin\theta_o\right]\left[\hat{x}\cos\theta+\hat{z}\sin\theta\right]\\
&\vec{E}_{2\omega,ps}=\frac{3}{4}\alpha_d\sin^3\beta\cos^2\theta_i\sin3\phi\,\hat{y}\\
&\vec{E}_{2\omega,ss}=-\frac{3}{4}\alpha_d\sin^3\beta\sin3\phi\,\hat{y}\\
\end{aligned}
\end{equation}
with a slightly different result obtained for the pp-polarization:
\begin{eqnarray}\nonumber
\vec{E}_{2\omega,pp}=\left(\alpha_u+3\alpha_d\cos^3\beta)\sin^2\theta_i\sin\theta_o-3\alpha_d\cos\beta\cos\theta_i\cos\theta_o\sin^2\beta\sin\theta_i\right.\\
+\frac{3}{4}\alpha_d\cos^2\theta_i\sin^2\beta(\cos\theta_o\cos3\phi\sin\beta+2\cos\beta\sin\theta_o)
\label{eq:23}
\end{eqnarray}

By setting an arbitrary DC offsets which might be related to experimental noise, the model showed good agreement with the experiment, as is evident from the reproducable symmetry pattern with correct azimuthal position. In the simulation we have -for the sake of simplicity- set the values for the hyperpolarizabilities equal to unity and use for the SiO-Si interface incoming and outgoing angle of $29.5^\circ$ (via snellius). The 3-fold $sp$ symmetry (green line) is perfectly matching the experimental result as well as the $ps$-polarization blue line) with the correct relative peak difference between them ($30^\circ$).

The intensity profiles can be easily explained as an effect of anharmonic charge radiation along the bond direction. One can easily check this  by calculating the field produced by the individual bond contribution. For example, the $6$-fold symmetry from the $ss$-polarization can be explained by each down bond contributing twice to the $s$-fundamental driving field when rotated by $360^\circ$. However, it has to be stated that for the $pp$-polarization SBHM gives a 3-fold symmetry and is in contradiction with the 6-fold symmetry given by experiment. We attribute this difference due to possible bulk dipole contribution because inside the bulk the incident angle is closer to the normal, and here SBHM also predicts a sixfold symmetry. This bulk dipole is still controversial and requires symmetry breaking in the form of a decaying field (absorption). Nevertheless, our simulation has shown that SHG measurement with SBHM can be used to investigate bond orientation at the surface of centrosymmetric materials  in a more  simple fashion without going to the complicated susceptibility tensor analysis (e.g. group theory).

\section{Summary}
Using the simplified bond hyperpolarizability model we show that the reflected linear intensity for all 4-polarization modes from a  Si(111) structure is independent of the azimuthal rotation on the $x$-$y$ plane. For the second harmonic intensity, SBHM gives a good agreement with experiment, sucessfully predicting the correct azimuthal profiles and symmetries of the surfaces except for the $pp$-polarization which we believe requires SBHM to be extended to cope with bulk effects. This shows that SBHM can be used as a simple model in predicting the bond orientation of a Si(111) wafer at the surface/interface and reconfirms second harmonic generation measurements as a sensitive non destructive method to investigate surface structures in material with inversion symmetry. 

\emph{Acknowledegements}: The authors would like to thank financial support by the Austrian Federal Ministry of  Economy, the Austrian Family and Youth and the Austrian National Foundation for Research, Technology and Development.

\end{document}